\documentclass[preprint,showpacs,preprintnumbers,amsmath,amssymb]{revtex4}

\usepackage{graphicx}

\usepackage{dcolumn}
\usepackage{bm}
\usepackage{color}

\begin{document}

\title{Ferromagnetic Behavior of High Purity ZnO Nanoparticles.}
\author{R. Escudero}
\email[Author to whom correspondence should be addressed. Email address:]
{escu@servidor.unam.mx}

\author{R. Escamilla}
\affiliation{Instituto de Investigaciones en Materiales, Universidad Nacional Aut\'{o}noma de M\'{e}xico. A. Postal 70-360. M\'{e}xico, D.F.
04510 MEXICO.}

\date{\today}

\begin{abstract}ZnO nanoparticles with Wurtzite structure were  prepared by chemical methods at low temperature 
in aqueous solution. Nanoparticles are in the  range from about 10 to 30 nm.  Ferromagnetic properties were  observed  from 2 K to room temperature and above.  
 Magnetization  vs temperature, M(T), and isothermal measurements M(H) were determined. The coercive field clearly shows 
ferromagnetism  above room temperature. An  exchange bias was observed, and we related this behavior at a core shell structure presented in the samples.
 The chemical synthesis, structure, defects in the bulk related to oxygen vacancies are the main factors for the observed magnetic behavior.  

\end{abstract}


\maketitle

\section{Introduction}
Recently Zinc oxide, ZnO has been a material that has attracted an intense interest  for its  
possible application in distinct technological fields; spintronics, magnetic semiconductors,
catalysts, sensors, field emission devices, solar cells, etc \cite{Ozgur}. This compound  is  a band gap semiconducting material 
 with a direct band gap of 3.37 eV, and electronic concentration only about $10^{12}$ to $10^{14}$ $cm^{-3}$. In bulk material Zinc atoms occupy the special position 2(b) with coordinates (1/3, 2/3, 0), oxygen also occupy special position 2(b) with coordinates (1/3, 2/3, u), where  u= a/c$(3/8)^{1/2}$ = 0.3817, see Kisi, et al. \cite{kisi}.  Figure 1 presents the hexagonal structure characteristic of zinc-oxide. The ZnO wurzite structure also named as zincite mineral, has cell parameters  
$a=3.249$ \AA, and $c=5.206$ \AA, consistent with the standard JCPDS 36-1451 pattern card. 

\begin{figure}[btp]
\begin{center}
\includegraphics[scale=0.1]{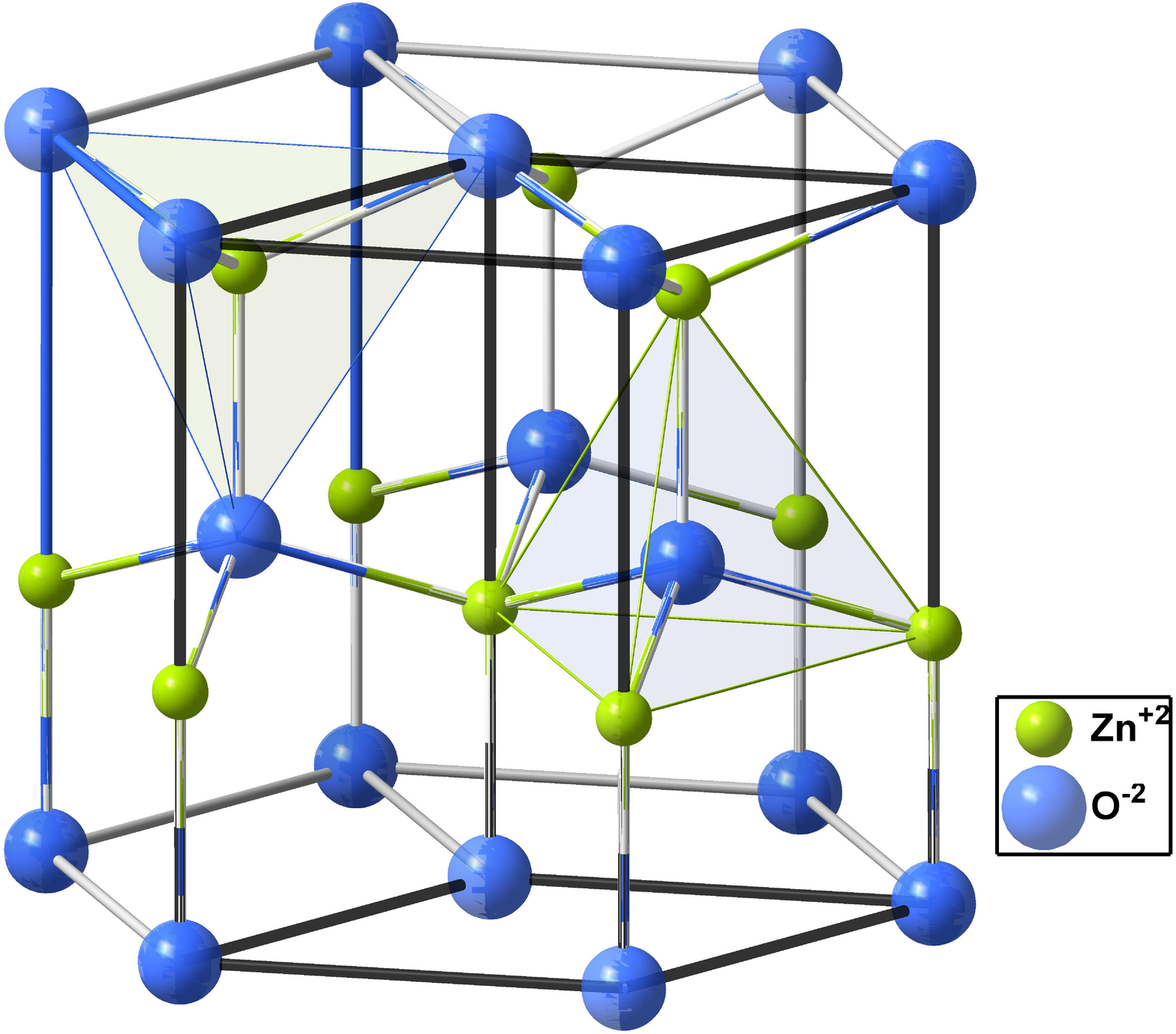}
\end{center}
\caption{(Color online) Wurzite crystalline structure of ZnO. In this figure the two  tetrahedra show a different configuration for the
 anions and cations: One shows  a cation as the central atom whereas the another shows the anion surrounded by four cations at the corners. This coordination 
forms  a typical  sp$^3$ covalent bonding.} 
\label{fig1}
\end{figure}

In addition, to the studies in bulk crystalline materials,
nanoparticles of this compound with different morphologies,  have been  turned important  for the reason that 
the electronic characteristics may change at the nanometric scale \cite{Yanhe,zhang,sundaresan,sundaresan2,nguyen}. These  changes are  mainly related to the ratio volumen/area, $V/A$
which may give different physical characteristics and behaviors. Recently studies of native points defects in ZnO bulk by Anderson, et al. \cite{Anderson}, describe the mechanism of defects over the electronic properties. In compounds at the nanometric scale the ratio $V/A$ changes the physical and chemical properties
  making posible diferent electronic mechanisms. Many workers in the field of nanomaterials  have used different procedures  for the preparation of this material; using  varied physical and chemical techniques 
to obtain the synthesis of this compound \cite{Yanhe,Hulan}. In this work we prepared  the nanostructured material using a  Sol-gel method as described by Hui, et al. \cite{Xiangyang}.
The  obtained nanostructures present interesting and new  magnetic characteristics some already reported by other authors, \cite{nguyen,sundaresan,sundaresan2}.  
In this paper we present our studies concerned with structural defects in ZnO nano-material and  the magnetic ordering found. Our observation shows that ZnO in sizes from about 10 to 30 nm  presents ferromagnetic ordering up to room temperature, which can be related to the absence or vacancies of oxygen or Zn atoms \cite{Kay}, XPS 
measurements confirm the assumption that vacancies are an important factor.

\section{Experimental Details}
Nanoparticles of ZnO were prepared using the  sol gel method. Different amounts of di-hidrated zinc acetate  Zn(CH$_3$COO)$_2$,  
(Aldrich 99.99 \%) was used as precursor diluted in de-ionized water at pH = 7 and 25 $^0$C, ammonium hydroxide (NH$_4$OH)was added slowly  as a dispersive medium 
according to the following reaction: 

         Zn(CH$_3$COO)$_2$ + 2NH$_4$OH   ----->    Zn(OH)$_2$ + 2CH$_3$COO$^-$ + 2NH$^+$$_4$.

 The resulting gel was dried at 60 $^0$C in air for a  period of time (30 hrs) and posteriorly  annealed in air,  at maximum temperature about 200 - 300 $^0$C
depending of the precursor amount to obtain ZnO,  the reaction was:  
							Zn(OH)$_2$  -----> ZnO +H$_2$O.

Different powder samples of ZnO  were prepared with distinct pH´s values and precursor amounts \cite{ismai}.
 In this work particularly we will describe mainly the resulting study done on three samples called: TO-1, TO-2, and TO-8A. 
The samples were characterized and analyzed with X-ray diffraction, see Fig. 2. There    
 we present the X-ray diffraction patterns for three  ZnO nanoparticle samples.  The panels show the X-ray nanoparticles with average sizes from  
 about 30 to 13 $nm$, the size was determined with Scherer formula and using a gaussian fit, the center of the gaussian was considered
 the average size. All the diffraction peaks were appropriated indexed, and corresponding  with  the hexagonal
 Wurtzite ZnO structure. 

\begin{figure}[btp]
\begin{center}
\includegraphics[scale=0.4]{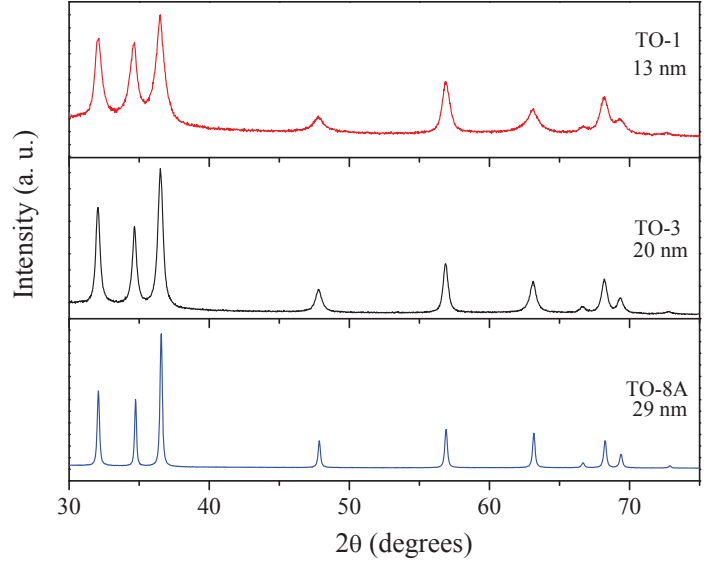}
\end{center}
\caption{(Color online) X-ray powder diffraction data for three nanoparticle samples. The  panels display the X-ray data 
size,  from top to bottom: 13, 20, and 29  $nm$. the powder spectra for all the nanopowders correspond with a pure 
structure without impurities. Data correspond with the wurzite hexagonal
 structure, and  cell parameters as mentioned before.}
\label{fig2}
\end{figure}

Phase identification of the samples was performed with  X-ray diffractometer Brucker D8 using Cu-K$_\alpha$ radiation and a Ni filter. 
Intensities were measured at room temperature in steps of 0.025 degrees, from  2$\theta$ range 6 to  130 degrees. The crystallographic phases were identified by comparison
 with the X-ray patterns of the JCPDS database. The parameters were refined with a Rietveld-fit program; Rietica  v 1.71 with 
multi-phase capability \cite{howard}. The structural parameters for three ZnO samples ( TO-1, TO-3, and TO-8A) are shown in Table 1, 
there are concentrated all  characteristics including parameters and  Rp, Re, and Rwp data related to the Rietveld fitting. 
Chemical analysis was carried out by X-ray photoelectron spectroscopy (XPS). The analysis was performed  using a VG Microtech ESCA 2000 Multilab UHV system, with an Al $K_\alpha$ X-ray source, h$\nu$ = 1486.6 eV, operated at 15 kV and 20 mA beam, 
and CLAM4 MCD analyzer.

\begin{figure}[btp]
\begin{center}
\includegraphics[scale=0.4]{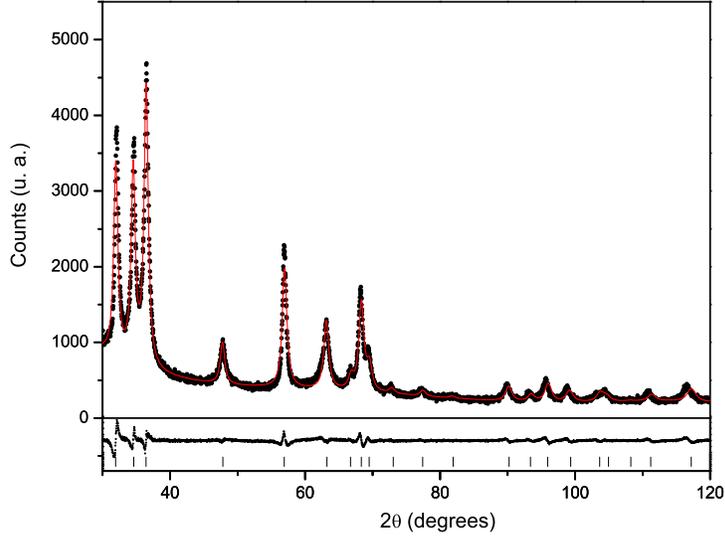}
\end{center}
\caption{(Color online) Rietveld refining data for a 20 $nm$ sample corresponding  with wurzite hexagonal
 structure. The  crystalline structure in different nanosamples is related 
 with deficiencies of oxygen or metal atoms into the structure.}
\label{fig3}
\end{figure}

  In order to perform the analyisis, the surface of the pellets were etched with Ar$^{+}$ for 20 min with 4.5 kV at 0.3 $\mu$A mm$^{-2}$. 
The XPS spectrum was obtained at 55 degrees respect to the normal surface in
 a constant pass energy mode (CAE), at E$_0$ = 50 and 20 eV for survey and high resolution narrow scan, respectively. The atomic relative sensitivity
 factor (RSF) reported by Scofield was corrected by the transmission function of the analyzer \cite{scofield} and by the reference material ZnO. The peak
 positions were referenced to the background silver 3d$_{5/2}$ photopeak at 368.21 eV, having a FWHM of 1.00 eV, and C-1s hydrocarbon groups in 284.50 eV
 central peak position. The XPS spectra were fitted with the program SDP v 4.1 \cite{SDP}. 

\begin{figure}[btp]
\begin{center}
\includegraphics[scale=0.7]{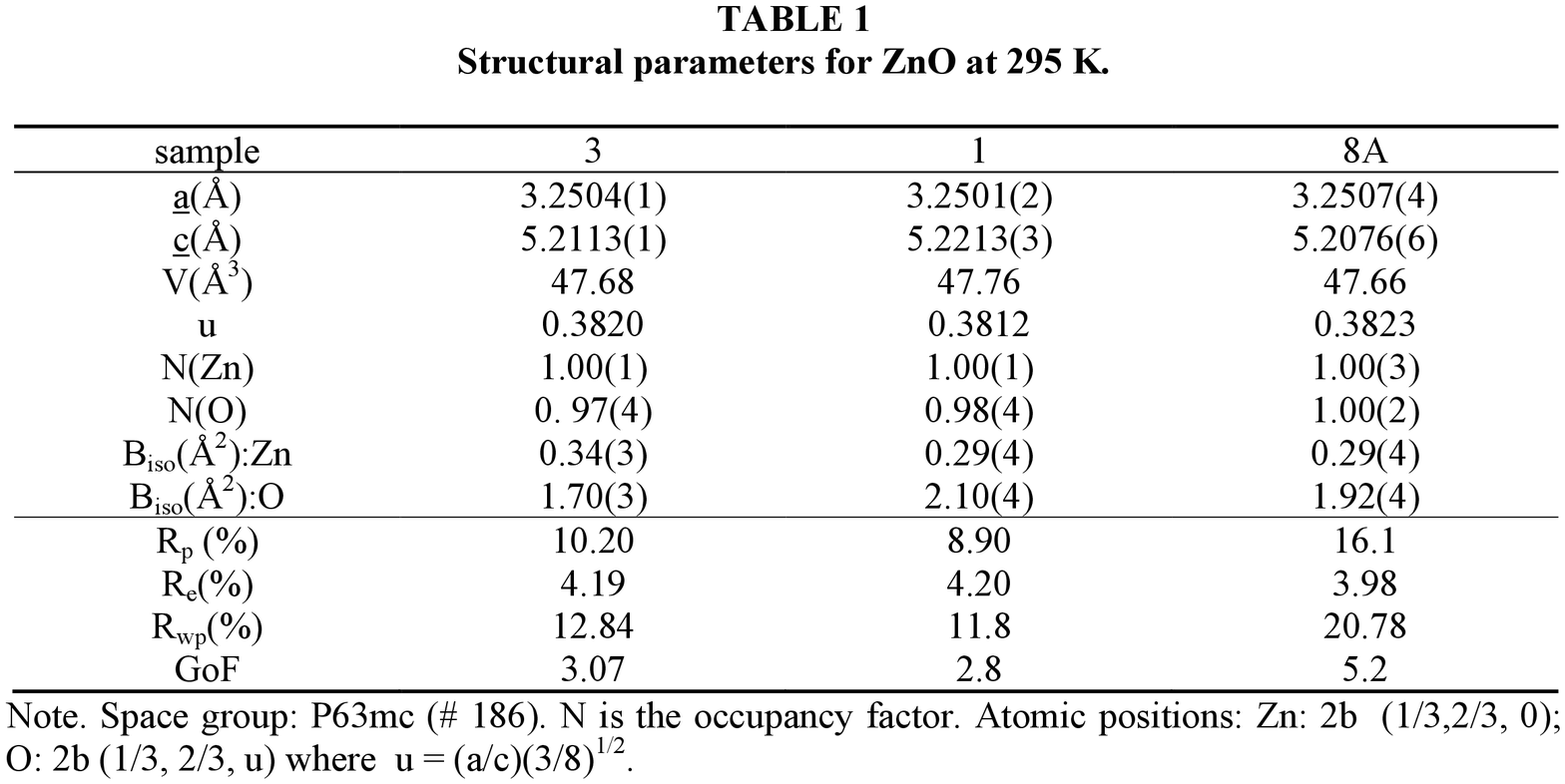}
\end{center}
\label{fig1a}
\end{figure}

Transmision electron micrographs (TEM) were obtained with  a JEOL FEG 2010 FASTEM analytical 
microscope.
Results of sinthesis,  X-ray, and TEM  show that nanoparticles with size below $15$ nm present a light grey color, 
whereas other with size above $25$ nm are with normal white coloration. 
TEM photographies of the nanostructured samples shown in Fig. 4 display the morphology and size of these nanoparticles. It is interesting to note 
 a core shell structure completely surrounded some of the nanoparticles, delimiting the bulk of the sample from the surface. This feature was  also observed by Debjani et al. \cite{Debjani}. 
We assume that these characteristic structural morphologies are related from the thermodynamical point of view, to one of the forms to decrease the ground
 state energy of the system formed by nanoparticles, and is specificaly  applicable  to our ZnO nanoparticles. 
As we will explain further, the  magnetic characteristic are affected by this shell structure, the effects mainly observed in an exchange bias feature. 
The  isothermal  magnetic measurements, particularly with the exchange bias in our  measurements will be displayed in a latter figure (Fig. 9). 
those results).

\begin{figure}[btp]
\begin{center}
\includegraphics[scale=0.5]{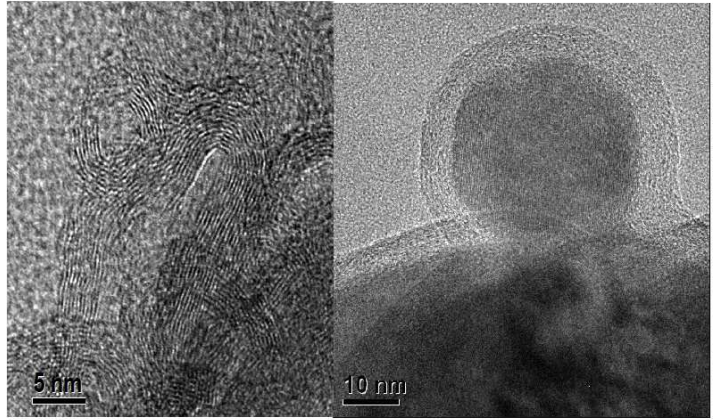}
\end{center}\caption{Transmision electron microscopy (TEM) of  ZnO nanoparticles. The TEM images display
the size of typical nanoparticles. Note the core shell structure observed. We related this core shell structure that presents the great majority of the particles as responsibles 
of the magnetic behavior, particularly to the exchange bias observed, and shown in figure 9.}
\label{fig4}
\end{figure}

In order to examine the stoichiometry of the compound, we analyzed the poly-crystalline samples by x-ray photoelectron spectroscopy (XPS). 
Fig. 5 shows the XPS survey spectra after Ar+ etching for the three samples. No extra peak corresponding to any magnetic impurities
 other than Zn and O atoms were  observed, as shown in this figure.

\begin{figure}[btp]
\begin{center}
\includegraphics[scale=0.4]{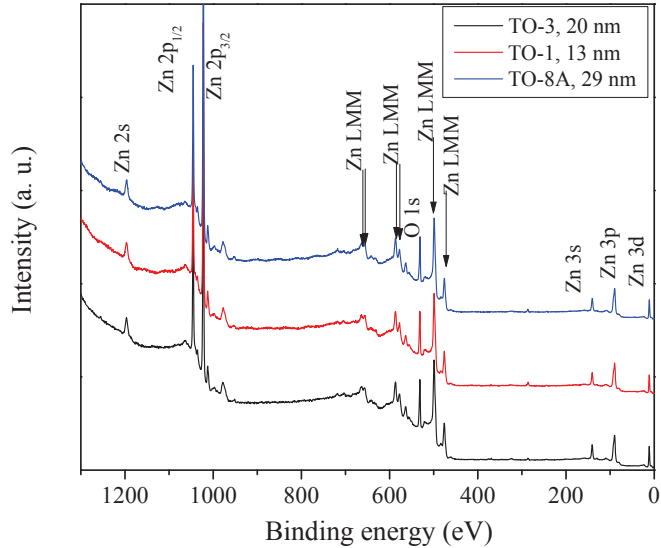}
\end{center}
\caption{(Color online)XPS surveys spectra after Ar$^{+}$ etching for polycristalline samples TO-1, TO-3, and TO-8A.  No extra peaks corresponding to any magnetic 
impurity were observed. Only are present Zn and O atoms without  other metal transition impurities.}
\label{fig5}
\end{figure}

\begin{figure}[btp]
\begin{center}
\includegraphics[scale=0.5]{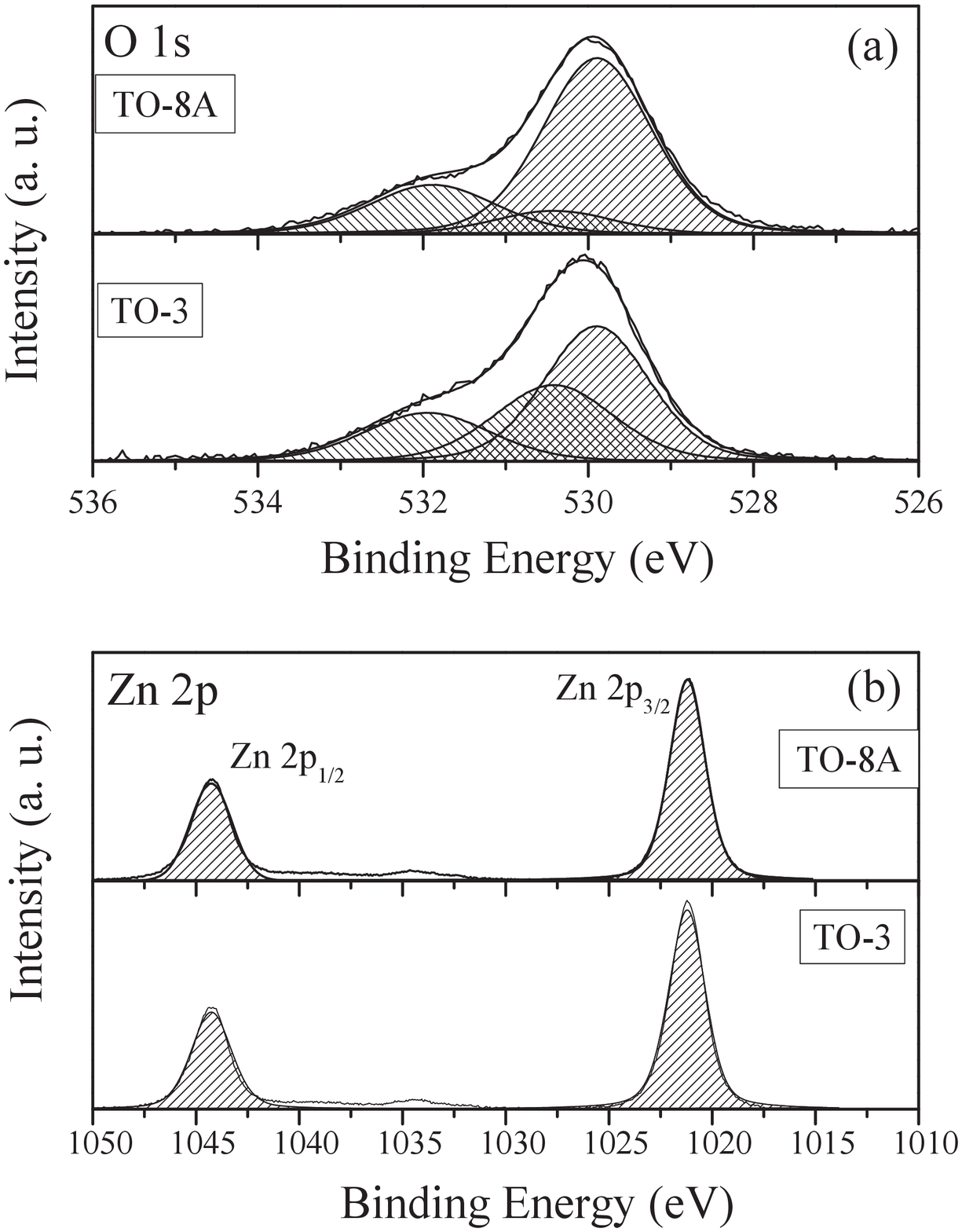}
\end{center}
\caption{High resolution XPS spectra of the a) O-1s, and b) Zn-2p core levels for two poly-cristalline samples  TO-3, TO-8A. the detailed shape is fitted 
assuming the contribution for three   components  belonging to two different chemical states of ZnO bulk sample. The lower binding energy
component  at 529.90 eV is attributted  to the structure of  hexagonal Zn$^{+2}$. The medium and higher binding energy components  
at 530.42 and 531.96 eV are assigned to oxygen deficiencies regions in the ZnO.}
\label{fig6}
\end{figure}

In Fig. 6(a) is shown the deconvolution of the XPS spectra in the O-1s region for TO-3 and TO-8A samples. The asymmetric O-1s peak in the surface was 
fitted by three Lorentzian-Gaussian components, centered at 529.90, 530.42, and 531.96 eV, respectively, as displayed in Fig. 6(a). The three fitted binding energy peaks 
approximate the results of Chen et al., and Wang et al. \cite{Chen, Wang}. Chen et al. attributed the peak on the low binding energy side of the O-1s spectrum 
to the O$^{2-}$ ions on the wurtzite structure of the hexagonal Zn$^{2+}$ ion array, surrounded by zinc atoms with the full supplement of nearest-neighbor O$^{2-}$ 
ions. Accordingly, this peak of the O-1s spectrum can be attributed to the Zn-O bonds. The higher binding energy at 531.96 eV is usually attributed to chemisorbed or dissociated oxygen or OH species on the surface of the ZnO thin film, such as CO$_3$, adsorbed H$_2$O or adsorbed O$_2$ \cite{Chen, Major}. The component at the medium binding energy of the O-1s peak may be associated with O$^{2-}$ ions that exist in oxygen-deficient regions within the ZnO matrix. Thus as a result, changes in the 
intensity of this component may be in connection with the variations in the concentration of the oxygen vacancies \cite{SZo}. The intensity of the peak localized at 
529.90 eV exceeds those localized at 530.42 and 531.96 eV, indicating the strong Zn-O bonding in ZnO. 
In order to determine the sample compositions, the atomic concentration was calculated by XPS using the survey spectra and the value of RSF for O-1s and Zn-2p: 2.95, and 9, respectively. For the three smples studied; TO-8A, TO-3 and TO-1, the $\%$ of atomic oxygen was determined as 0.757, 0.736 and 0.715 respectively. 
In Fig. 6(b) are shown the deconvolution of the XPS spectra in the Zn-2p region for the mentioned samples. The Zn-2p spectrum shows a doublet whose binding energies are 1021.21 and 1044.26 eV and can be identified as the lines Zn-2p$_{3/2}$ and Zn-2p$_{1/2}$, respectively. The binding energy differences between the two lines is 22.4 eV, 
which is well lying within the standard reference value of ZnO \cite{wagner}. The values of binding energy and  binding energy difference  calculated from the XPS study show that Zn atoms are in Zn$^{2+}$ oxidation state. No metallic Zn with a binding energy of 1021.5 eV was observed \cite{Islam}. This confirms again that Zn is  in the Zn$^{2+}$ state. 
Furthermore, at this stage is not possible to determine  the presence of Zn interstitial defects from Zn-2p spectra, which may be established only by 
the Auger peak of Zn.

\subsection{Magnetic characterization}

Magnetic behavior was determined from low temperature, 2 K to room temperature. The characteristics were determined by using 
a MPMS Quantum Design Magnetometer. Magnetization versus Temperature characteristics
 M(T) were performed in magnetic field intensities of 1 kOe and 50 kOe, and in two typical 
modes of measurement, in order to determine probably irreversible effects.
 In Zero Field Cooling  (ZFC) the sample
is first cooled to the minimum accesible temperature, once  in thermal equilibrium the magnetic field is applied and the measurement
is initiated increasing  the temperature. In the field cooling mode (FC) the measurement is performed starting at the 
maximum temperature with the magnetic field applied and  decreasing the temperature. With those two modes, 
the behavior  related to hysteretic or irreversible processes can readily be observed. In addition to these measurement, we also 
performed measurements at the two mentioned different magnetic fields, this in order to observe possible metamagnetic effects; in all  measurements
neither metamagnetic or irreversibility processes were observed. In Fig. 7, it is   displayed a tipical susceptibility characteristic as function of temperature,
 for two different size particle samples measured at two distinct fields. In general we  observed  similar  magnetic behavior for  all  samples, with small
variation of the Curie constant from about 0.05 to  0.09 molK/cm$^3$, with small changes of the Weiss temperature,  to about  250 to 326 K. 
The small Curie values extracted of the  susceptibility are quite representative of a weak ferromagnetic behavior.  Fig. 7 displays 
both sets of data in two different samples. Top panel also shows  the inverse susceptibility $\chi^{-1}(T)$ fitting  very well to a Curie Weiss
 law from high temperature to below 50 K. The resulting Curie constant is; C = 0.0795 molKcm$^3$, and  Weiss temperature $\theta_\omega = 270$ K.
 The bottom panel shows  $\chi(T)$, and the right axis  the number of Bohr magnetons, $\mu_{eff}$,  at room temperature is about 0.55 BM, value similar to all samples.

\begin{figure}[btp]
\begin{center}
\includegraphics[scale=0.7]{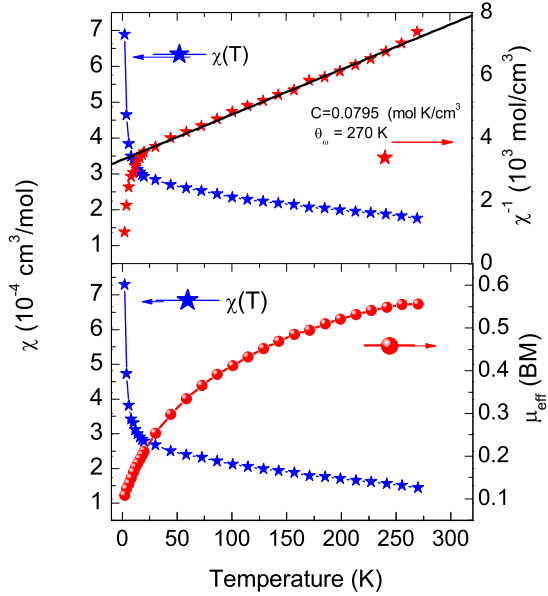}
\end{center}
\caption{(Color online) Top panel  presents the magnetic susceptibility as function of temperature, $\chi(T)$, left  axis is $\chi(T)$ in terms of cm$^3$/mol. The characteristic behavior 
is at high temperature a simple  Curie Weiss behavior. The right  axis shows the inverse susceptibility, full line indicates the fitting   values 
of C = 0.0795 molKcm$^3$, and $\theta_\omega = 270$ K.
At the lower panel  we also present another typical data for $\chi(T)$, with Curie  constant C = 0.055 molKcm$^3$, the number of Bohr magnetons  is shown at the right with 
room temperature value of only 0.55 $BM$.}
\label{fig7}
\end{figure}

\begin{figure}[btp]
\begin{center}
\includegraphics[scale=0.3]{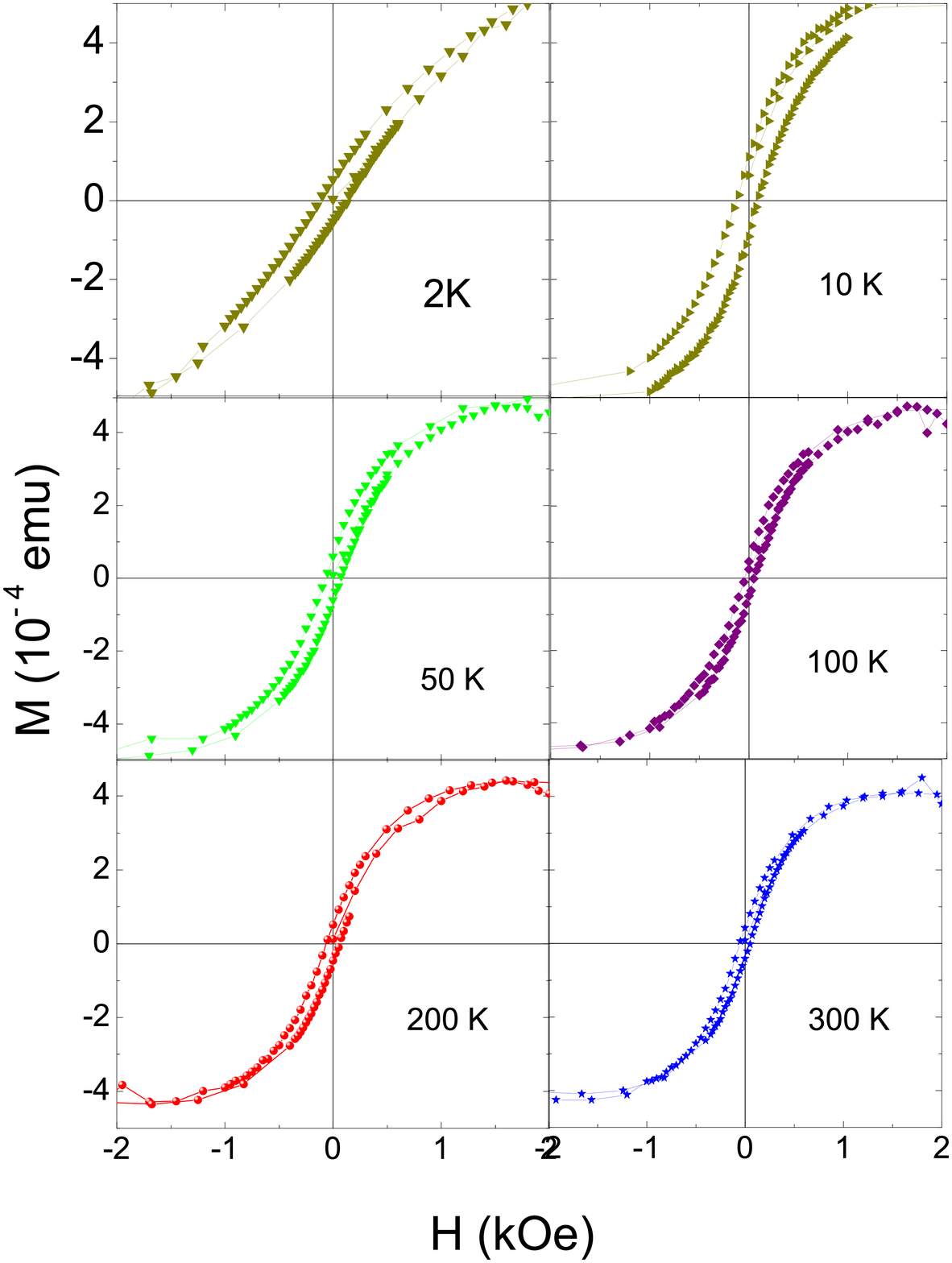}
\end{center}
\caption{(Color online) Isothermal measurements, $M-H$ at six different temperatures. These show a coercive 
field that is big at low temperature and decreases as the temperature increases. In addition it display an
 exchange bias quite  different at low temperatures. Data were measured from 2 to 300 K.} 
\label{fig8}
\end{figure}

In order to have better insight about the magnetic behavior of the ZnO nanosamples, we performed  isothermal magnetization measurements, M(H),  determined 
at different temperatures to characterize the resulting coercive 
field. These measurements were from 2 K to above room temperature.  Fig. 8  displays  panels with isothermal M(H) at six different temperatures. 
All temperatures show an anomalous variation of the  coercive field as displayed in Fig. 9. Several interesting characteristics can be noted;   
at low temperature the coercive field presents an anomalous peak around  5 K. Measurements taken at 2 K present a small value that increases, and reaches a maximum at 
about 5 K. After this temperature the coercive field decreases up to 200 K. Above 200 K it seems that another anomalous change ocurrs at 250 K followed for a smooth decrease, 
but persistent to above room temperature. We have plotted in Fig. 9 the positive and negative parts of the coercive field. Clearly is  noted a small but different coercive field in the  TO-3 and TO-8A samples, and also in the negative and positive parts. This exchange bias we speculate may be  related to two possible magnetic orderings
 ocurring  in the samples; one existing on  surface and other in  bulk. The core shell charateristic of our samples are surely  related to  this behavior \cite{schuller} and also explain our  observations. However we do not have a complete explanation for the two observed anomalous peaks of the coercive field at low temperature. 
It is important to stress about the purity of the samples. Particularly, we did not find detect any  magnetic impurities.       
The determined exchange bias differences, respect to the negative and positive parts plotted in Fig. 9 are  small but quite measurable.   
Accordingly this effect may be due to two different magnetic orderings  in the bulk and on surface of the samples. 
Thus, this explains our results of the exchange bias.

\begin{figure}[btp]
\begin{center}
\includegraphics[scale=0.3]{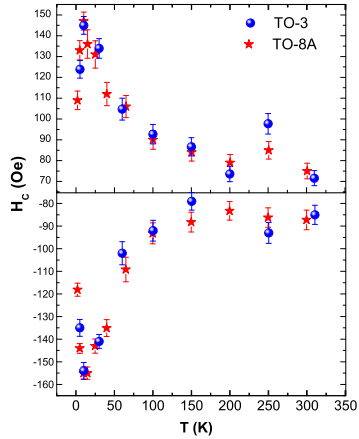}
\end{center}
\caption{(Color online) Plot of the Coercive field displaying a exchange bias, more clearly observed  at low temperature. This general exchange bias behavior is  clearly 
the influence of the core shell morphology of the nanoparticles.}
\label{fig9}
\end{figure}

\section{Conclusions}
In this experimental study of ZnO nanoparticles prepared using sol gel method we demonstrate that ferromagnetism
 exist up to room temperature. Rietveld refining shows that the cell parameters are almost similar to bulk samples 
and studies of the XPS show the presence of oxygen vacancies,  around 25 - 29 \% . That difference in  structure, is conclusive and related  with the magnetic behavior. 
 Also we observed an exchange  bias in measurements related to the coercive field, this effect might be  clearly related to a  core shell structure as 
observed in our samples and by other workers, this structural shell definitive affects the magnetic order in the nanoparticles.

\begin{acknowledgments}
We thank F. Silvar for Helium provisions, L. Huerta for XPS measurements, L. Rendon, for high resolution TEM measurements, L. Diazbarriga, for the help in the conclusion of this study, and O. Garcia, A. Fernandez for initial help in initial method of sample preparation.
\end{acknowledgments}


\begin{thebibliography}{apsrev}
\bibitem{Ozgur} U. Ozgur, Ya. I. Alivov, C. Liu, A. Teke, M. A. Reshchikov, S. Dogan, V. Avrutin, S. J. Cho, and H. Morkoc. J. Appl. Phys. 98, 041301 (2005).
\bibitem{kisi}Erich H. Kisi and Margaret M. Elcombe. Acta Cryst. C45, 1867 (1989).

\bibitem{Yanhe} Yanhe Xiao, Liang Li, Yan Li, Ming Fang and Lide Zhang. Nanotechnology 16, no. 6, 671 (2005).

\bibitem{zhang} Hui Zhang, Deren Yang, Xiangyan Ma, Yujie Ji, Jin Xu, Duanlin Que. Nanotechnology 15, 622 (2004).
\bibitem{sundaresan} A. Sundaresan, R. Bhargavi, N. Rangarajan, U. Siddesh.  C. N. R. Rao. Physical Review B 74, 161306 (R) (2006).
\bibitem{sundaresan2} A. Sundaresan, C. N. R. Rao. Solid State Communications. 149, 1197 (2009).

\bibitem{nguyen} Nguyen Hoa Hong, Joe Sakai, Virginie Brizé. J. Phys:. Condens. Matter 19, 036219 (2007).


\bibitem{Anderson} Anderson Janotti and Chris G. Van de Walle. Phys. Rev. B 76, 165202 (2007). 

\bibitem{Hulan} Hulan Zhou, Zhuang Li. Materials Chemistry and Physics 89, 326 (2005). 

\bibitem{Xiangyang} Xiangyang Ma, Hui Zhang, Yujie Ji, Jin Xu, Deren Yang. Materials Letters 59, Issue 27, 3393 (2005) 

\bibitem{Kay} Kay Potzger, and Shengqiang Zhou. Phys. Status Solidi B 246, No.6, 1147 (2009).

\bibitem{ismai}A. A. Ismail, A. El-Midany, E. A. Abdel-Aal, H. El. Shall. Materials Letters 59, 1924 (2005).

\bibitem{howard}C. J. Howard, B. A. Hunter, DE.A. J. Swinkels. Rietica, IUCR Powder Diffraction 22, 21 (1997).
\bibitem {scofield} J. H. Scofield. J. Electron Spectrosc. 8, 129 (1976).
\bibitem{SDP} SDP, v4.1 (32bit) Copyright 2004, XPS International, LLC. Compiled Jan. (2004).
\bibitem{Debjani} Debjani Karmakar, S. K. Mandal, R. M. Kadam, P. L. Paulose, A. K. Rajarajan, T. K. Nath, A. K. Das, I. Dasgupta and G. P. Das. Phys. Rev. B 75, 144404 (2007).

\bibitem{Chen} M. Chen, X. Wang, Y. H. Yu, Z. L. Pei, X. D. Bai, C. Sun, R. F. Huang, L. S. Wen. Appl. Surf. Sci. 158, 134 (2000).
\bibitem{Wang} Z. G. Wang, X. T. Zu, S. Zhu, L. M. Wang. Phys. E 35, 199 (2006).
\bibitem{Major}S. Major, S. Kumar, M. Bhatnagar, K. L. Chopra. Appl. Phys. Lett. 49, 394 (1986).

\bibitem{SZo}T. Ször´enyi, L. D. Laude, I. Bert´oti, Z. K´antor, Z. Geretovszky. J. Appl. Phys. 78, 6211 (1995).

\bibitem{wagner} C. D. Wagner, W. M. Riggs, et al. in: Handbook of X-ray photoelectron Spectroscopy, Perkin Elmer, Eden Praire. (1979).

\bibitem{Islam}M. N. Islam, T. B. Ghosh, K. L. Chopra, H. N. Acharya. Thin Solid Films 280, 20 (1996).
 

\bibitem{schuller}J. Nogues, I. K. Schuller. Journal of Magnetism and Magnetic Materials 192, 203 (1999).

\end{thebibliography}
\end{document}